# Instrumento livre para medidas de movimento
*(Free instrument for movement measure)*

Norberto Peña[2], Bruno Cecílio Credidio[1], Lorena Peixoto Nogueira Rodriguez Martinez Salles Corrêa[1], Lucas Gabriel Souza França[1], Marcelo do Vale Cunha[1], Marcos Cavalcanti de Sousa[1], João Paulo Bomfim Cruz Vieira[1] e José Garcia Vivas Miranda[1]

[1] *Laboratório de Física Nuclear Aplicada – Instituto de Física da Universidade Federal da Bahia*
[2] *Departamento de Fisioterapia - Faculdade de Ciências da Saúde da Universidade Federal da Bahia.*

Resumo

Este trabalho apresenta a validação de uma ferramenta computacional gratuita, que serve para obtenção de medidas contínuas de objetos em movimento. O software utiliza técnicas de visão computacional, reconhecimento de padrões e fluxo óptico, para viabilizar o rastreamento de objetos em vídeos, gerando dados de trajetória, velocidade, aceleração e movimento angular. O programa foi aplicado no rastreamento da esfera de um pêndulo aproximadamente simples. A metodologia utilizada para a validação toma como base a comparação dos valores medidos pelo programa, bem como os valores teóricos esperados segundo o modelo do pêndulo simples. O experimento se adequou ao método, pois foi construído respeitando os limites lineares do oscilador harmônico, minimizando as perdas de energia por atrito e tornando-o o mais ideal possível. Os resultados indicam que a ferramenta é sensível e precisa. Desvios menores do que um milímetro na medida da trajetória garantem a aplicabilidade do software em física, tanto em projetos de pesquisa quanto em tópicos de ensino.
Palavras-chave: ensino de física, visão computacional, pêndulo simples.

Abstract

This paper presents the validation of a computational tool that serves to obtain continuous measurements of moving objects. The software uses techniques of computer vision, pattern recognition and optical flow, to enable tracking of objects in videos, generating data trajectory, velocity, acceleration and angular movement. The program was applied to track a ball around a simple pendulum. The methodology used to validate it, taking as a basis to compare the values measured by the program, as well as the theoretical values expected according to the model of a simple pendulum. The experiment is appropriate to the method because it was built within the limits of the linear harmonic oscillator and energy losses due to friction had been minimized, making it the most ideal possible. The results indicate that the tool is sensitive and accurate. Deviations of less than a millimeter to the extent of the trajectory, ensures the applicability of the software on physics, whether in research or in teaching topics.

1. Introdução

O estudo do movimento no ensino de física ou em laboratórios de pesquisa orbita quase sempre entre dois extremos: medidas de fácil obtenção e medidas com alta precisão e contínuas. As primeiras são imprecisas, necessitam de poucos recursos (normalmente se baseiam em cronômetros e réguas) e apresentam valores médios como resultado, como é o caso comum de experimentos em laboratório do ensino médio. No outro extremo aparecem medidas com alta precisão e contínuas, que utilizam tecnologias exigentes de recursos superiores e que são de difícil operação em relação às anteriores. O pesquisador/educador que busca medidas contínuas do movimento não encontra opções de instrumentos com características intermediárias.

Um importante exemplo dessa dualidade ocorre em instrumentos de análise do movimento humano. Para a compreensão de sua complexidade e caracterização é essencial uma análise objetiva da motricidade. Esta costuma ser realizada de forma clínica através de testes epidemiológicos [1, 2] ou utilizando tecnologias como plataformas de força [3], vídeo-análise [4], sistemas optoelectrônicos [5], eletro-goniômetros [6], giroscópios [7], eletromiógrafos [8, 9, 10] e acelerômetros [11]. Estes estudos utilizam instrumentos sofisticados e de alto custo, que analisam eventos relacionados com o deslocamento angular, reação das forças do chão e o controle motor em indivíduos e em populações. Aplicações simples em pesquisa ou educação se tornam inviáveis.

O crescimento constante da capacidade de processamento dos computadores, aliado à adoção de novos paradigmas de desenvolvimento colaborativos, permitem a construção de programas de alta complexidade com baixo custo. Em tempos atuais é factível, utilizando um computador pessoal, construir ferramentas de seguimento de fluxo em sequência de imagens capazes de prever a trajetória de pontos no espaço utilizando uma simples câmera de vídeo digital. As bibliotecas de tratamento de vídeo e visão computacional OpenCv [12] e OpenGinga [13] são exemplos desse avanço, pois colaborativa e gratuitamente disponibilizam diversas funções de tratamento de vídeo otimizadas, que permitem a elaboração de programas sofisticados de análise de movimento.

O objetivo geral deste trabalho é apresentar uma nova ferramenta de análise do movimento construída com base na biblioteca OpenCv, através da utilização de técnicas de visão computacional capazes de preencher a lacuna tecnológica com medidas precisas e de baixo custo. Também é apresentada uma metodologia de validação de instrumentos de medição de trajetórias com base em tecnologias de visão computacional, além de uma aplicação para o estudo do movimento humano em eventos de sentar e levantar.

## 2. O Instrumento

Trata-se de um medidor de parâmetros mecânicos do movimento (trajetória, velocidade e aceleração) com base em algoritmos de visão computacional aplicados a vídeos de objetos em movimento. O instrumental necessário consiste em uma câmera digital e o programa livre desenvolvido, o CVMob [14]. As especificações mínimas da câmera dependerão da velocidade com que ocorre o evento que se queira estudar. Para o fenômeno observado neste trabalho – um pêndulo simples – foi necessária uma câmera com resolução de filmagem de 640x480 pixels com uma taxa de captura de 30 quadros por segundo (fps). A partir dessas especificações foi possível uma descrição completa do movimento com medições da trajetória, velocidade e aceleração.

O CVMob é um software livre que utiliza técnicas de Visão Computacional com análise do fluxo de pixels em vídeos, para localização e acompanhamento de padrões de imagens. O CVMob foi desenvolvido no Instituto de Física da Universidade Federal da Bahia em linguagem C++, utilizando o *framework* QT4 e a biblioteca de visão computacional OpenCV.

Um experimento de validação foi realizado para identificar os limites de medidas do instrumento e pode ser generalizado para qualquer situação em que se utilizem técnicas de visão computacional como base de medida.

## 3. Métodos de validação

A implementação de uma nova ferramenta de medida requer uma imprescindível análise criteriosa de seu desempenho. De acordo com Ribeiro e colaboradores [15] existem dois parâmetros cruciais no desempenho de qualquer técnica analítica: a qualidade das medidas instrumentais e a confiabilidade estatística dos cálculos envolvidos no seu processamento. O estabelecimento dos limites de precisão desses parâmetros utiliza comparações com base em eventos teóricos conhecidos, leis e/ou estimativas anteriores sobre o fenômeno, o que assegura a aplicabilidade e o alcance da análise. Este conjunto de procedimentos é conhecido como validação.

O experimento utilizado na validação das medidas consistiu na determinação da trajetória e da velocidade máxima de um pêndulo solto a diferentes alturas. Na Figura 1A apresentamos uma ilustração esquemática. O pêndulo foi elevado a alturas conhecidas para estimação dos parâmetros. A equação de ajuste da projeção da trajetória no eixo x foi:

$$X(t) = Xo + A\,sen(\pi\frac{t-tc}{w}) \quad (1)$$

Na qual *Xo* e *tc* são parâmetros de ajuste que transladam a curva teórica para os valores iniciais medidos pelo CvMob, *A* é a amplitude e *w* o meio período. Para a validação do instrumento como estimador da velocidade utilizou-se a expressão teórica da velocidade máxima atingida em um pêndulo simples,

$$v_{max} = \sqrt{2gh} \qquad (2)$$

Onde *h* representa a altura inicial do pêndulo e *g,* a aceleração da gravidade. A altura inicial *h* foi medida com o auxílio de um laser que, alinhado com a linha do pêndulo e os marcadores de altura no anteparo, faz com que o erro na configuração inicial seja menor que o diâmetro da projeção do feixe no anteparo (~2 mm). O pêndulo foi solto rompendo-se o fio de nylon que o prende a um suporte.

Como a liberação do pêndulo pode se tornar outra fonte de erros, o fio de nylon foi queimado para provocar o rompimento. Isto garantiu que a velocidade inicial da esfera massiva presa ao pêndulo fosse zero, uma vez que não existiu interação direta do experimentador com ela. A massa do pêndulo utilizado foi muito maior que a massa do fio de nylon, fazendo com que o aumento de massa do sistema devido ao nylon seja desprezível. Uma vez solto, o movimento foi filmado por dois períodos consecutivos, o que produz aproximadamente de 260 pontos medidos de trajetória, de acordo com as especificações da câmera e pêndulo.

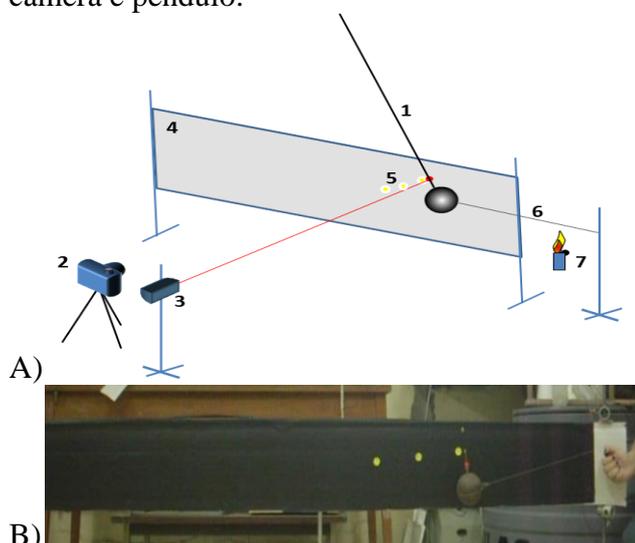

**Figura 1 – em A, esquema ilustrativo da montagem usada na validação do instrumento 1) pêndulo, 2) câmera digital, 3) laser, 4) anteparo, 5) marcadores de posicionamento da altura inicial, 6) linha de nylon e 7) isqueiro. Em B foto do anteparo com marcadores e pêndulo usados.**

As especificações do pêndulo foram: uma esfera de ferro com 7,52 cm de diâmetro equatorial, 670g de massa presa por um fio com 3,00m. A câmera utilizada foi uma Casio FX com uma resolução de filmagem de 640x480 pixels capturados a 30fps, a partir de uma lente com distância focal de 35mm.

Na Figura 2, há um exemplo de tela do programa CvMob em um procedimento de medida. Os dados são capturados e exportados para uma planilha de cálculo de onde os valores da trajetória e da velocidade máxima são extraídos.

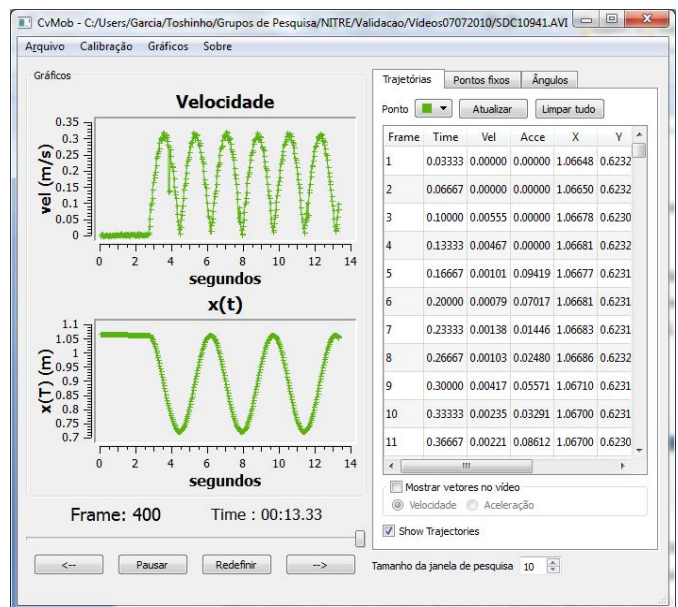

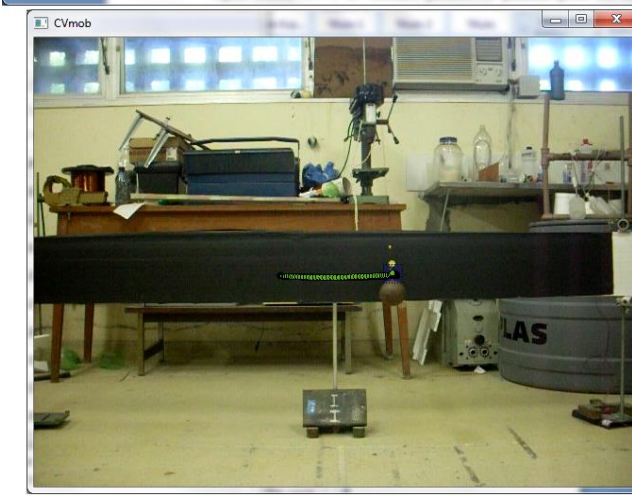

**Figura 2 – Janelas do programa CvMob em um procedimento de medida.**

A precisão na aferição das medidas foi estimada em dois níveis: repetibilidade e precisão intermediária [15]. Foram utilizados três valores de altura inicial (baixo, médio e alto) para determinar a faixa de velocidade. Para aferir a repetibilidade foram executadas 27 duplicatas, 9 para cada faixa de velocidade. Para aferir a precisão intermediária foram executadas 24 duplicatas, ou seja, 8 duplicatas por faixa de velocidade[1]. As aferições foram realizadas em dias diferentes, cada vez com uma equipe distinta, somando um total de 51 vídeos registrados. O diâmetro do pêndulo foi utilizado como escala de calibração nos vídeos.

### 4. Resultados da validação

Os vídeos gerados em cada repetição x aferição foram analisados com o programa CvMob e as medidas de trajetória e velocidade máxima comparadas com modelos teóricos.

#### 4.1. Trajetória do pêndulo

A medida da trajetória, além do respectivo ajuste teórico de um dos vídeos estão na Figura 3. Nota-se que a curva teórica passa por todos os pontos medidos. O ajuste dos dados ao modelo teórico (1) por mínimos quadrados resulta num coeficiente de correlação de Pearson de $R^2=0,99995$ e a soma dos quadrados dos resíduos $\Upsilon=2,066 \times 10^{-4}$m, o que leva a um desvio padrão do ajuste (DPA) de $8,97 \times 10^{-4}$m (~0.9mm). O mesmo procedimento foi executado para todos os 51 vídeos, sendo obtido o valor do DPA para cada um deles. O valor máximo do DPA encontrado no conjunto de 51 vídeos analisados foi de $8,80 \times 10^{-3}$m.

Na Tabela 1 são apresentados os resultados dos ajustes da trajetória para três valores de altura inicial nas duas aferições. Esses valores representam o erro padrão mínimo, o padrão médio e o padrão máximo, obtidos a partir da análise dos 51 vídeos.

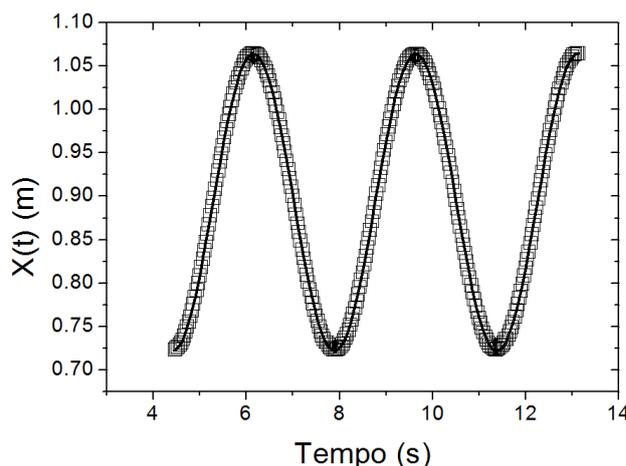

**Figura 3** – Ajuste do modelo teórico senoidal às medidas obtidas com o CvMob. Os quadrados representam as medidas; a linha sólida, o ajuste.

Na Tabela 1 são indicados os valores da altura inicial *Ho*, do período *2w*, do DPA e da aceleração da gravidade estimada mediante (3)

$$g = \frac{\pi^2 l}{w^2} \qquad (3)$$

onde *l* é o comprimento do fio que vai do ponto de apoio do pêndulo até o centro da esfera (3,00m para 1ª aferição e 2,9m para 2ª).

**Tabela 1** – Valores dos ajustes do modelo teórico.

|  | $Ho^*$(m) | $2w^{**}$(s) | $DPA^+$(m) | g (m/s²) |
|---|---|---|---|---|
| 1ª Aferição | 0,414 | 3,4836(2) | 0,0013 | 9,757(6) |
| | 0,428 | 3,4842(4) | 0,0026 | 9,754(6) |
| | 0,440 | 3,4858(5) | 0,0039 | 9,745(6) |
| 2ª Aferição | 0,534 | 3,4221(5) | 0,0027 | 9,775(6) |
| | 0,571 | 3,4264(5) | 0,0052 | 9,751(6) |
| | 0,603 | 3,4269(8) | 0,0088 | 9,748(6) |

* Altura inicial, ** Período, + Erro padrão

Os resultados apresentados na Tabela 1 indicam uma pequena tendência de aumento do DPA com o aumento da altura inicial. Nota-se também que os erros para a segunda aferição foram maiores do que na primeira. Isso se deve às condições de luminosidade em que foram capturados os vídeos na segunda aferição. Na ocasião existia menos luz e as filmagens apresentaram um efeito de granularidade maior do que na primeira.

---

[1] com exceção da altura inicial média da primeira aferição, que por problemas na captura de dois vídeos foi avaliada apenas 8 vezes.

Para um melhor entendimento da origem das flutuações nas medidas, foi realizada uma avaliação de como os resíduos oscilam em torno da curva teórica ajustada. No gráfico da Figura 4, pode-se notar que os resíduos calculados a partir do vídeo sem nenhum tratamento (normal), apresentam dois pontos de descontinuidade em 3,8 e 7,7 segundos. Observando detalhadamente a filmagem, nota-se que no método de captura da câmera existe um mecanismo de interpolação de quadros. Para manter a taxa de captura constante a câmera inclui, em alguns momentos, uma cópia do quadro anterior. É ocasionada uma translação da curva e consequentemente um aumento no erro do ajuste. No exemplo da Figura 4, o vídeo foi tratado eliminando-se os dois quadros copiados e repetiu-se o procedimento de medida e estimativa do erro. Com essa alteração, as descontinuidades desapareceram e o erro diminuiu de 0,0039m para 0,0030m. Mesmo conhecendo a origem de parte do erro optou-se por manter o método sem tratamento, uma vez que o procedimento de busca e eliminação de quadros repetidos é complexo. Além disso, o objetivo proposto foi avaliar a precisão do método e do instrumento com todos os seus artefatos.

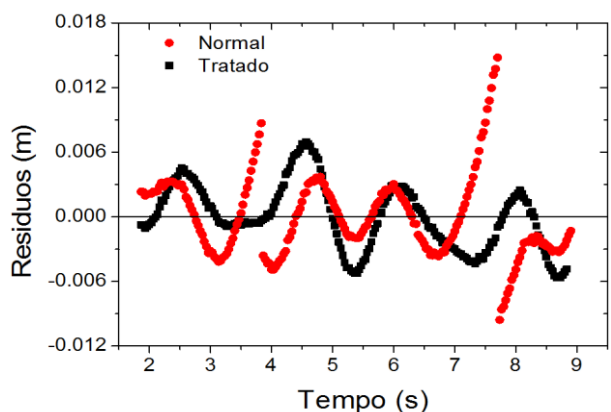

**Figura 4 – Resíduos do ajuste ao modelo no vídeo 10970 com e sem tratamento.**

Para avaliar a influência da flutuação na luminosidade da sala no algoritmo de busca para rastreamento do ponto, mediu-se as flutuações oriundas da trajetória de um ponto fixo no anteparo. Para todos os vídeos capturados, o desvio padrão das medidas não ultrapassou 0,0003m no conjunto de medidas da primeira aferição e 0,0004m na segunda, cerca de 10 vezes menor que o encontrado para pontos em movimento (Tabela 1). Desta forma, acredita-se que os erros estiveram relacionados aos seguintes fatores: atrito; deformação da imagem devido à lente da câmera; ruídos no circuito da câmera e às perdas devido ao processo de compactação da imagem.

### 4.2. Velocidade máxima do pêndulo

O segundo critério de validação do instrumento foi a comparação entre a velocidade máxima esperada pelo modelo teórico e a medida no instrumento.

Na Tabela 2, temos as médias dos valores medidos com seus respectivos valores teóricos e o desvio padrão experimental.

**Tabela 2 – Comparação entre as medidas de velocidade máxima (dados em m/s).**

| | Teórico | Média | Desvio Padrão |
|---|---|---|---|
| 1ª Aferição | 0,28(1) | 0,31 | 0,01 |
| | 0,59(1) | 0,58 | 0,02 |
| | 0,77(1) | 0,78 | 0,03 |
| 2ª Aferição | 0,49(1) | 0,52 | 0,05 |
| | 0,97(1) | 0,98 | 0,03 |
| | 1,25(1) | 1,30 | 0,04 |

Todos os desvios em relação à média apresentaram uma distribuição significativamente normal no teste de Kolmogorov Smirnov. Além disso, observou-se a comparação entre os valores médios com os teóricos, que evidenciam uma perfeita compatibilização, ao serem considerados os respectivos erros e desvios padrões.

Um importante fator no processo de validação é a linearidade das medidas. Os gráficos das Figuras 5 e 6 mostram a correlação linear entre os valores medidos e os teóricos; desta, resultam correlações de $R^2=0,98$ para primeira aferição e $R^2=0,9843$ para a segunda com inclinações próximas a 1 (0,95 e 1,02) e interseções próximas a zero (0,04 e 0,02).

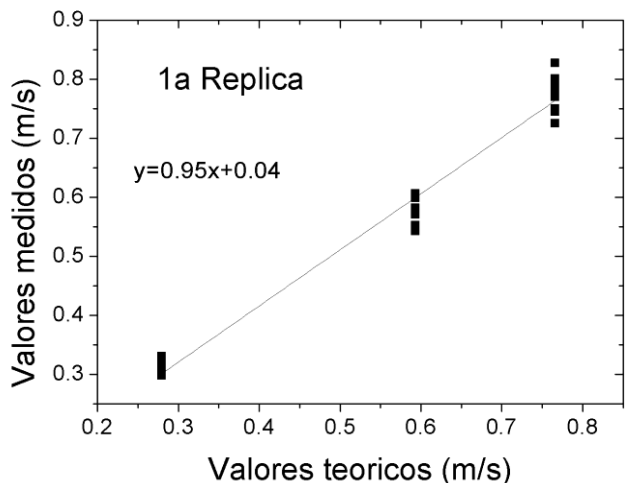

**Figura 5** – Análise da linearidade das medidas para o conjunto de medidas da 1ª aferição.

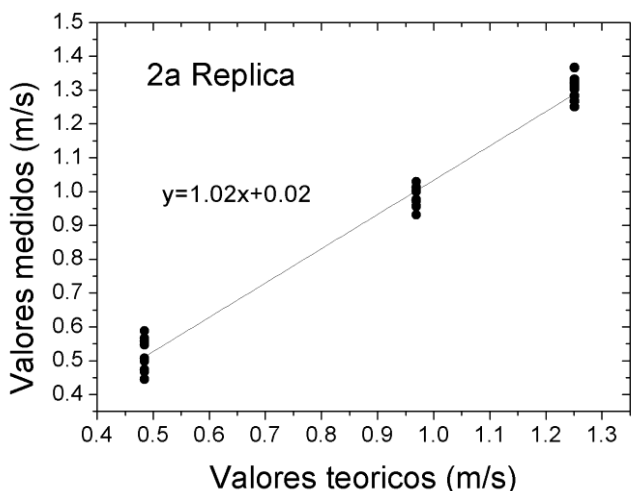

**Figura 6** – Análise da linearidade das medidas para o conjunto de medidas da 2ª aferição.

A sensibilidade do método foi definida pelos limites de detecção (LD) e de quantificação (LQ) [15]. O LD é a menor variação de velocidade que pode ser detectada pela técnica instrumental, enquanto o LQ é a mais baixa velocidade que pode ser quantificada. Os parâmetros da curva analítica LD e LQ podem ser observados na Tabela 3.

**Tabela 3.** Limite de detecção (LD) e de quantificação (LQ) da velocidade máxima estimados para o método com 95% de confiança. Valores em m/s.

| Aferições | LD   | LQ   |
|-----------|------|------|
| 1a        | 0,14 | 0,21 |
| 2a        | 0,19 | 0,28 |

## 5. Conclusões

Uma margem de segurança de 5% abaixo dos limites regulatórios foi alcançada para todas as velocidades para as estimativas do LQ realizadas com 95% e 99% de confiança. Todas as faixas de velocidade apresentaram LD e LQ abaixo de 5% dos limites regulatórios para um nível de 99,9% de confiança. Os valores de LD foram sempre menores que os valores de LQ, conforme o esperado.

Após a análise dos resultados obtidos, verifica-se que a utilização da tecnologia computacional validada, constitui um meio facilitador da análise do movimento, a partir de dados objetivos e precisos. Trata-se de um projeto que utiliza tecnologia de fácil transferência, uma vez que o software utilizado é de licença Livre. A partir destes resultados, se espera a popularização do uso do CvMob, o incentivo às pesquisas sobre motricidade humana ou animal e uma contribuição direta à democratização das tecnologias.

## 6. Referências